# Topologically tuned terahertz confinement in a nonlinear photonic chip

Jiayi Wang[1,*], Shiqi Xia[1,*], Ride Wang[2,*], Ruobin Ma[1], Yao Lu[1], Xinzheng Zhang[1,3]‡, Daohong Song[1,3], Qiang Wu[1,3], Roberto Morandotti[4], Jingjun Xu[1,3]‡, Zhigang Chen[1,3,5]‡

1The MOE Key Laboratory of Weak-Light Nonlinear Photonics, TEDA Institute of Applied Physics and School of Physics, Nankai University, Tianjin 300457, China

2Innovation Laboratory of Terahertz Biophysics, National Innovation Institute of Defense Technology, Beijing 100071, China

3 Collaborative Innovation Center of Extreme Optics, Shanxi University, Taiyuan, Shanxi 030006, China

4 INRS-EMT, 1650 Blvd. Lionel-Boulet, Varennes, Quebec J3X 1S2, Canada

5 Department of Physics and Astronomy, San Francisco State University, San Francisco, California 94132, USA

[*]These authors contributed equally to this work.

‡ *zxz@nankai.edu.cn, jjxu@nankai.edu.cn, zgchen@nankai.edu.cn*

**Abstract:** Compact terahertz (THz) functional devices are greatly sought after for high-speed wireless communication, biochemical sensing, and non-destructive inspection. However, conventional devices to generate and guide THz waves are afflicted with diffraction loss and disorder due to inevitable fabrication defects. Here, based on the topological protection of electromagnetic waves, we demonstrate nonlinear generation and topologically tuned confinement of THz waves in a judiciously-patterned lithium niobate chip forming a wedge-shaped Su-Schrieffer-Heeger lattice. Experimentally measured band structures provide direct visualization of the generated THz waves in momentum space, and their robustness to chiral perturbation is also analyzed and compared between topologically trivial and nontrivial regimes. Such chip-scale control of THz waves may bring about new possibilities for THz integrated topological circuits, promising for advanced photonic applications.

The surge of interest and development in reliable terahertz (THz) technology are driven by the high demand for applications such as in wireless communications[1,2], signal processing[3-5], biosensing[6] and non-destructive evaluation[7]. However, the lack of integrated functional devices in THz range including THz emitters has hampered their applications. In addition, it has always been a challenge to guide and manipulate THz waves due to the critical features of THz spectrum and unavoidable losses from fabrication defects and environmental conditions. Therefore, there have been tremendous efforts in exploring different designs and approaches for THz sources and integrated THz devices using different platforms, including for example metamaterials[3,5] and nonlinear metasurfaces[8], surface plasmonic waves[9], nonlinear wave mixing in ionic crystals[10], and time-domain integration of THz pulses[11].

Topological photonic systems, gifted with appealing edge and interface modes immune to disorder and impurities[12-17], have presented excellent potential for many applications such as in topological insulator lasers[18-20]. However, limited by material platforms and characterization methods, most research on topological photonics thus far has been focused on either microwave or optical wave regimes. Recently, the concept of topological phase of light has been explored for implementation in THz waveguides and circuits, and for the development of THz communications[21-23]. In particular, by building up a domain wall between two structures with opposite valley-Chern numbers, it has been shown that THz waves can transmit through sharp bends without significant losses due to the topological protection on the valley-Hall edge states[22,24]. These topological photonic structures are expected to be highly beneficial for compact and robust THz functional devices.

In this work, we propose and demonstrate a scheme for topologically-tuned THz-wave confinement upon nonlinear generation, all in a single lithium niobate (LN) photonic chip. Such a scheme relies on a judiciously designed photonic microstructure – a one-dimensional Su-Schrieffer-Heeger (SSH) lattice[25,26] consisting of LN waveguide stripes with wedge-shaped air gaps to undergo topologically trivial to nontrivial transitions. The structure is fabricated via the femtosecond-laser writing technique, with a varying interface topological defect in the center (Figs. 1a, 1b). Using a pump-probe experiment, we directly observe the nonlinear THz-waves with tunable confinement along the chip with respect to the variation of the photonic structure. We obtain the band structures by mapping the signals measured from time-resolved spectroscopy into momentum space, where topological nontrivial states are clearly identified. The existence and robustness of edge states to chiral perturbation are theoretically analyzed which shows excellent agreement with experiments. Our results provide a clear evidence of THz-wave confinement with decreased decay due to topological protection.

The simplest technique for THz-wave generation is based on the optical rectification (OR)[27-28] excited by femtosecond laser pulses in nonlinear crystals[29-31]. Such a nonlinear process normally produces a quasi-DC signal under continuous wave (CW) excitation written as $P^{(2)}(0) = \chi^{(2)}(0;\omega,-\omega)E(\omega)E^*(\omega)$, where $P$, $\chi^{(2)}(0;\omega,-\omega)$ and $E(\omega)$ represent the polarization, the second-order nonlinear susceptibility of samples and the driving electric field, respectively. When a femtosecond laser pulse is employed, the THz radiation occurs due to the nonlinear polarization induced by the intense laser pulse. Assuming mainly electronic contribution, the nonlinear polarization is mathematically described by[28]

$$P(\Omega) = \int_{\omega-\Delta\omega/2}^{\omega+\Delta\omega/2} \chi^{(2)}(\Omega;\omega'+\Omega,-\omega')E_p(\omega'+\Omega)E_p^*(\omega')d\omega', \qquad (1)$$

where $\omega$ and $\Delta\omega$ are the central frequency and the spectral width of the pump laser pulse, respectively, and $\Omega$ represents the frequency of the generated THz waves. $E_p(\omega'+\Omega)$ and $E_p^*(\omega')$ are the electric fields of the pump laser corresponding to different frequency components. The electric field $E(\Omega)$ of generated THz waves is proportional to the nonlinear polarization $P(\Omega)$. Over the past decades, numerous techniques have been developed to enhance the THz generation efficiency, achieve narrow THz bandwidth, and decrease the THz-wave decay in LN crystals[30-32]. For instance, it has been shown recently that giant enhancement of optical nonlinearity can be achieved at THz-frequency by stimulated phonon-polaritons in LN crystals pumped by a femtosecond laser pulse, where ionic contribution to the light-matter interaction becomes significant[10]. While the underlying mechanisms merit further investigation, it is experimentally evident that tunable THz pulses can be generated in nonlinear LN crystals with ultrashort laser pulses in the range from a few tenths of THz to a few THz[10,30,31].

Despite the rapid development in this field, new techniques are still desirable for THz-wave localization and confinement, as the THz waves typically spread out and decay quickly due to the loss and diffraction associated with the long-wavelength nature. Here we employ an SSH-type photonic lattice etched in an LN chip to achieve tunable topological THz-wave localization, with a frequency range of 0.1~0.8 THz in accord with the specific experimental parameters discussed later and detailed in the Supplementary Materials (SM)[33]. The SSH lattice serves as a prototypical topological model, and it has been widely demonstrated for instance in photonics[26,34] and plasmonics[35], highly tested for generation of entangled photon pairs[36], enhancement of nonlinear harmonic generation[37,38], and realization of topological lasing[39] and non-Hermitian topological states[40,41]. The SSH-type structure established in our experiment possesses a central interface topological defect[40,41,34], formed by fs-laser-writing of LN waveguide arrays with varying spacing along the chip (Fig. 1b)[33]. The distances between neighboring LN stripes are governed by:

$$d_1 = d_{10} - \delta d\, z/L, \quad d_2 = d_{20} + \delta d\, z/L \tag{2}$$

where $L = 6$ mm is the total length of the LN chip along $z$-axis with the distance $z$ measured from the top of the chip (Fig. 1b), and the dimer structure at the top is set by $d_{10} = 80$ μm and $d_{20} = 30$ μm. Here $d_1$ and $d_2$ represent the distances between adjacent stripes which determines the coupling coefficients $c_1$ and $c_2$, respectively: a larger distance $d$ leads to a weaker coupling $c$. Moreover, when $\delta d$ is set to nonzero (e.g., 50 μm in our case), we can have a "tunable" SSH-type lattice because the dimerization changes along $z$-axis. As seen in Fig. 1b, the defect is located at the center ($n = 0$), but it varies from a *long-long defect* (L-LD) (when $z < L/2$) to a trivial equidistance without defect (at $z = L/2$), and then to a *short-short defect* (S-SD) (when $z > L/2$), thereby we can achieve different topological phases of the SSH structure[34] in these three different regions (illustrated by different colors in Fig. 1a).

To better appreciate the difference in topological characteristics, we calculate the eigenvalues of the SSH lattice in different regions along $z$-axis and plot them in Fig. 2a. The yellow line in the middle denotes the equidistance as the topological phase transition point. Before this line, $z < L/2$ and $d_1 > d_2$, the lattice belongs to a topological structure with an L-LD at the interface (Fig. 1b). The topological property of the SSH model can be evaluated by their corresponding Zak phase[42], where in current situation the left part of L-LD illustrated on top of Fig. 1b owns a nonzero Zak phase but the right part contributes to a trivial Zak phase. As a result, nontrivial topological defect modes with eigenvalues residing in the middle of the gap are found with a frequency around 0.3 THz using above parameters (see red dots in Fig. 2a). The defect mode shown in Fig. 2b1 is located at the interface of two sub-lattices of the L-LD structure, where away from the defect the mode distributes only at the even-numbered lattice sites labeled in Fig. 1b with alternately opposite phase. Such an amplitude/phase distribution of the defect mode indicates the signature of nontrivial topology protected by the chiral symmetry of the SSH lattice[17,25,26]. Right at the transition point $z = L/2$, $d_1 = d_2 = 55$ μm according to Eq. 1, and the L-LD structure is transformed into a simple 1D equidistant lattice (see the yellow dashed line in Fig. 1b) with a closed gap in the band structure (see the yellow line in Fig. 2a). The structure corresponding to this point has no more dimerization and thus turns into a trivial periodic lattice. As such, the mode distributes across the whole lattice (no localized edge state) at 0.3 THz (Fig. 2b2). This point marks the occurrence of topological phase transition.

When $z > L/2$, this lattice structure turns into an S-SD regime and it becomes topologically nontrivial again, similar to the L-LD regime except now there are three LN stripes (lattice sites) closely spaced at the interface forming the "short-short defect" (see the bottom of Fig. 1b). Therefore, a topological defect mode also arises in the S-SD structure at around 0.3 THz (see green dots in Fig. 2a). However, the mode in this case is only distributed at odd-numbered lattice sites with alternately opposite phase, but no distribution on the central defect site ($n = 0$) (Fig. 2b4). Moreover, apart from the topological defect mode supported in the S-SD structure, there exists a trivial mode located around 0.42 THz. Such a trivial mode shows a peak power on the defect site but it also occupies all lattice sites (see blue dots in Fig. 2a and Fig. 2b3). Most importantly, as we shall show later, it has no more topological protection as compared to the nontrivial defect mode at around 0.3 THz. Due to such difference in the mode distribution, the excitation condition in the experiment determines if a trivial or a nontrivial mode would be excited in the S-SD end of the chip. Thus, along such a specially patterned LN lattice structure, the confinement of the THz waves generated by nonlinear excitation can be fine-tuned by scanning the pump beam through the structure to undergo a topological phase transition.

In order to realize the proposed THz manipulation, we perform a series of experiments with the typical pump-probe setup[10,33] to obtain dispersion curves of the structure at different $z$ positions (Fig. 3, first row) and related energy distributions of confined modes (Fig. 3, second row). In our experiments, as shown in Fig. 1a, a femtosecond pump beam (800 nm central wavelength, 120 fs pulse duration, 1kHz repetition rate) is cylindrically focused onto the center defect of the LN chip, thereby to generate THz waves via the nonlinear OR process. Due to the electro-optical effect of the LN crystal, lateral propagation of the generated THz waves (along *x*-direction in the chip plane) can be directly observed using a time-resolved imaging technique, thanks to the refractive index change induced by the THz waves as examined by a probe beam via the phase contrast imaging[10]. By varying the time delay between pump and probe pulses launched onto the LN chip, we can measure spatiotemporal evolutions of THz waves and then acquire their dispersion curves by performing a two-dimensional (2D) Fourier transform on the *x-t* diagrams (see more details in SM[33]).

Experimental results obtained with the LN sample shown in Fig. 1b are summarized in top two rows of Fig. 3. In the L-LD region ($z < 3$ mm), the topological defect mode appears around 0.3 THz, as can be clearly identified in the middle of the bandgap (Fig. 3a1). Close to the top ($z = 0$, location A in Fig 1b), most energy at this THz frequency is confined in the center defect with only weak sidelobes populating the even-numbered sites (Fig. 3a2). This strong localization characterizes the feature of topological defect mode in the L-LD structure, in good agreement with the calculation shown in Fig. 2b1. Closer towards the center, the localization becomes degraded since the SSH structure turns into the weakly nontrivial region (Fig. 3b). At the middle equidistance position ($z = 3$ mm, location C in Fig 1b), the structure undergoes a critical topological phase transition to a trivial lattice, so the THz waves spread into the bulk and the intensity localization disappears (Fig. 3c2). In this case, since the lattice period is halved (no dimerization), the range of Brillouin zone is doubled, and no defect mode can be identified in the spectrum (Fig. 3c1). Moving further to the S-SD region ($z > 3$ mm), we observe again strong localization but at a frequency around 0.42 THz. Close to the bottom ($z = 6$ mm, location E in Fig 1b), the localized THz waves at this frequency have intensity peaks in all lattice sites next to the center defect (Fig. 3e2), indicating that the trivial defect mode (as in Fig. 2b3) in the S-SD structure is excited. Even though the input Gaussian-like pump does not favor the excitation of nontrivial defect mode in this case, which has a dipole-like structure near the center defect (as in Fig. 2b4), the phase difference of THz waves spreading away from the center stripe caused by slightly tilting the pump beam leads to a minor trace of the topological state around 0.3 THz in experiment (see more details in SM[33]).

The above experimental results demonstrate clearly that the generated THz waves can be strongly confined near the center defect in both L-LD and S-SD regions (Fig. 3, second row), so long as it is far away from the transition point (marked in Fig. 2a). The significant difference between these two regions is that the localization of 0.3 THz in the L-LD (marked as A, B in Fig. 3) has topological protection, whereas the observed localization of 0.42 THz in the S-SD (marked as D, E in Fig. 3) results from trivial defect modes. These observations are further corroborated by numerical simulations presented in bottom two rows of Fig. 3, where Figs. 3(a4-e4) are obtained by performing 2D Fourier transform on the *x-t* diagrams in Figs. 3(a3-e3). There seems to be a difference between experimental and simulated results if one compares the top and the bottom rows. This is simply due to the instability of the laser which causes the divergency of the background at each moment, leading to unwelcome vertical lines in the spectra that also indicate the positions of the Brillouin zone edges (see more details in SM[33]). Apart from that, simulations show a good agreement with experimental results. Upon nonlinear excitation, the THz waves are generated and switched from a topological defect mode, to a bulk mode, and then to trivial defect mode, exhibiting tunable confinement along the LN chip.

Before closing, let us further analyze the distinction between topological nontrivial and trivial defect modes, using the following tight-binding SSH model with a center defect:

$$H = c_1\left(\sum_{n \in N_+} \xi_{n+1} a_n^\dagger a_{n+1} + \sum_{n \in N_-} \xi_{n-1} a_n^\dagger a_{n-1}\right) + c_2\left(\sum_{n \in N_+} \xi_{n+2} a_{n+1}^\dagger a_{n+2} + \sum_{n \in N_-} \xi_{n-2} a_{n-1}^\dagger a_{n-2}\right) + h.c.$$

$$N_+ = 2N, \ N_- = -2N, \ N = (0, 1, 2, 3 \ldots) \tag{3}$$

where $a_n$ ($a_n^\dagger$) is the annihilation (creation) operator in the $n$-th site of the lattice labeled in Fig. 1b, $\xi_n$ is the perturbation added on the coupling, and $c_1$ and $c_2$ describe the coupling coefficients between the LN stripe waveguides spaced by $d_1$ and $d_2$ in Fig. 1b, respectively. The coupling strength can be effectively tuned by changing the distance between neighboring stripes, where a smaller spacing leads to a larger coupling. When $\xi_n = 0$, $d_1 > d_2$ results in $c_1 < c_2$ in the L-LD case, but in the S-SD case it is $d_1 < d_2$ that results in $c_1 > c_2$. Since the salient characteristics of a topological mode are represented by the robustness against perturbations, we add chiral perturbations ($\xi_{-n} = \xi_n$) on all off-diagonal terms of the Hamiltonian in Eq. (3) (i.e., on all coupling coefficients without breaking the chiral symmetry[41]). To perform a quantitative analysis, we set $c_1 = 1, c_2 = 3$ for the L-LD, $c_1 = 3, c_2 = 1$ for the S-SD, and add 500 sets of perturbations with $\max(\xi_n) = 30\%$. For the L-LD case, the eigenvalue of a topological defect mode is robust and isolated from the bulk modes under perturbations (Fig. 4a1). In the S-SD structure, even though a topological mode is still robust, the trivial defect mode excited in Figs. 3(d, e) is severely affected by perturbations (Fig. 4b1). Direct comparison of numerically simulated *x-t* diagrams and corresponding spectra of the THz waves between the L-LD and the S-SD structures under perturbations also confirm the difference with respect to topological protection, as shown in Figs. 4(a2, a3) and Figs. 4(b2, b3). Therefore, topological structures can be employed to suppress or eliminate the scattering loss and decay of THz waves introduced by inevitable fabrication defects so long the chiral symmetry is preserved. As shown in Fig.

4c, even at a high perturbation (max($\xi_n$) = 60%) where the probability that a trivial defect mode couples into the bulk raises 50%, the topological nontrivial mode still preserves a robust confinement of THz waves (see more details in SM [33]).

In conclusion, we have demonstrated a scheme for nonlinear generation and topologically tuned terahertz confinement in a single photonic chip. The chip consists of an LN waveguide array with wedge-shaped air gaps fabricated with distinct (L-LD and S-SD) topological interface defects by fs-laser writing technique. Topological phase transition occurs in the middle of the LN chip, which features three regions of different topological characteristics for localization/delocalization of the nonlinearly generated THz waves. Theoretical analysis utilizing the tight-binding method and numerical simulations are in good agreement with experimental observations and further substantiate the distinctive features of confined THz topological states under chiral perturbations. This work offers a flexible and convenient way to tune the confinement as well as the topological properties of THz waves on demand, which may open an avenue towards the implementation of versatile, stable and compact THz photonic integrated circuits for various applications, including imaging, sixth-generation wireless communication, global environmental monitoring and high-speed data processing[2,7,22,43]. For further investigations, the study on different types of topological phenomena in the THz frequency range, such as the Weyl points[44-46], the quantum Hall effect[13,47,48], the Floquet topological phases[14,49,50] and parity-time symmetry in non-Hermitian topological systems[51-53], will bring about intriguing possibilities to manipulate THz waves and eventually contribute to the development of THz functional devices and the advancement of THz technologies.

**Acknowledgments**

This work was supported by the National Key Research and Development Program of China (2017YFA0303800, 2017YFA0305100), PCSIRT (IRT_13R29), Higher Education Discipline Innovation Project (B07013) and the National Natural Science Foundation of China (12134006, 91750204, 12074201, 11922408).

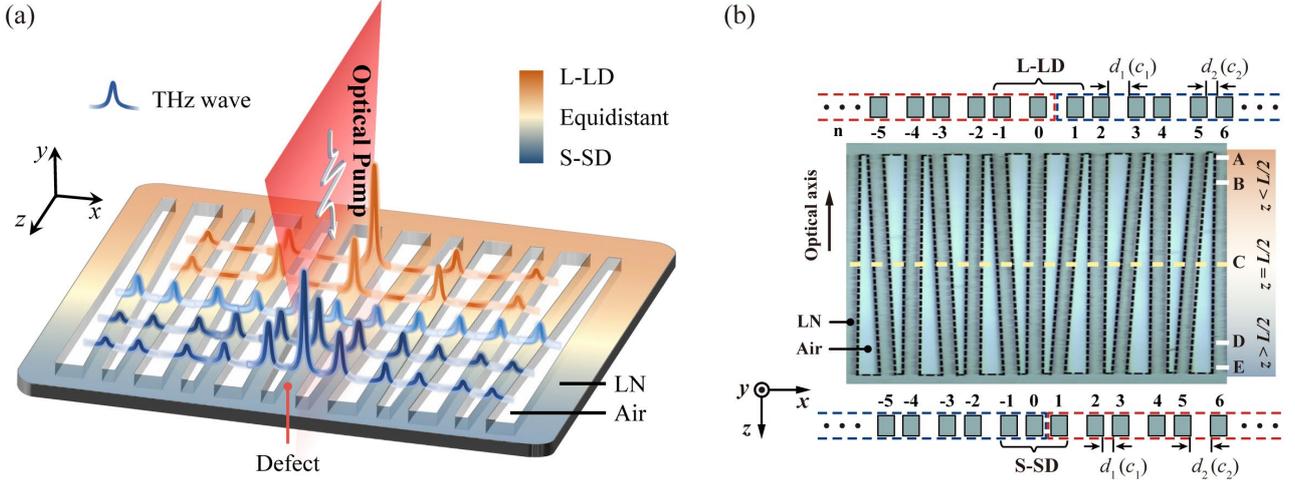

Fig. 1. Experimental realization of topologically controlled THz localization. (a) Illustration of nonlinear generation and confinement of THz-wave at different $z$ positions of an SSH-type microstructure. The LN structure undergoes a transition from L-LD, equidistance, to S-SD regions along the $+z$ axis, illustrated by colors shaded from orange into blue. (b) Microscope image of the LN array structure fabricated by fs-laser direct writing. The optical axis of the LN is along $z$ direction. The thickness of the LN chip is 50 μm in $y$-direction. The total length of the microstructure along $z$-direction is $L = 6$ mm. $d_1$ and $d_2$ are the spacings between neighboring LN stripes corresponding to coupling coefficients $c_1$ and $c_2$, respectively. At a dashed yellow line, $z = L/2$ and $d_1 = d_2 = 55$ μm, which leads to an equidistant structure. The SSH lattice contains an L-LD ($z < L/2$, $d_1 > d_2$, $c_1 < c_2$) above the line and an S-SD below the line ($z > L/2$, $d_1 < d_2$, $c_1 > c_2$). A, B, C, D and E denote the locations corresponding to $z = 0, L/8, L/2, 7L/8, L$. $n$ numbers the LN waveguides (lattice sites). Red and blue dashed lines mark the topological nontrivial and trivial parts of the SSH lattice.

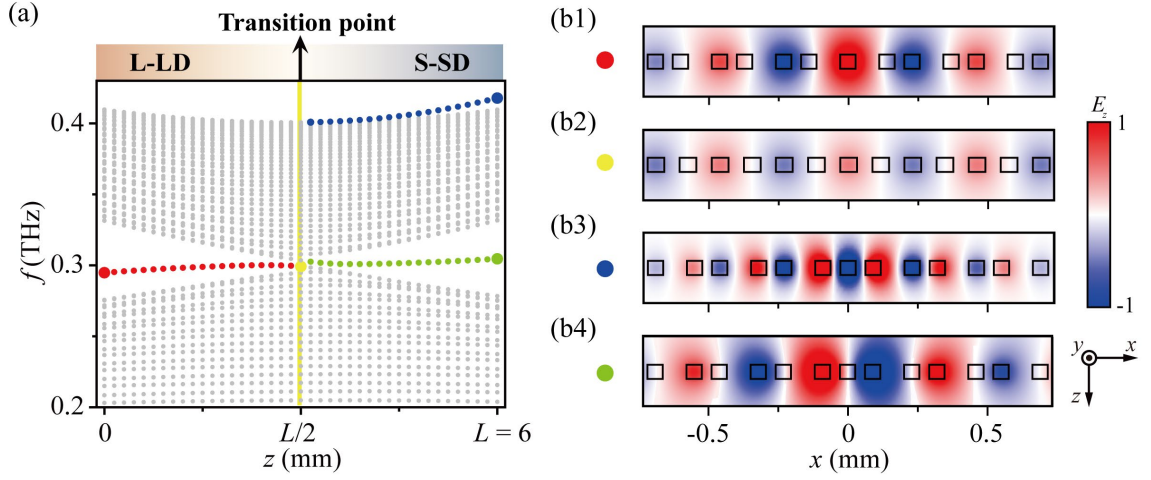

Fig. 2. Eigenvalues and representative eigenmode distributions in the SSH-type LN topological structure. (a) Calculated eigenvalue distribution of the microstructure along $z$-axis. The yellow line represents the equidistant structure at $z = L/2$ ($d_1 = d_2 = 55$ μm), which can be regarded as a phase transition point. The left side of yellow line ($z < L/2$) is the L-LD region, where topological defect modes are denoted by red dots. The right side ($z > L/2$) indicates the S-SD region, where topological nontrivial and trivial defect modes are marked by green and blue dots, respectively. Gray dots represent the bulk modes. (b1) Topological defect mode around 0.3 THz in the L-LD structure at $z = 0$. (b2) The mode around 0.3 THz in the equidistant structure at $z = L/2$. (b3, b4) Topological trivial mode around 0.42 THz (b3) and nontrivial mode around 0.3 THz (b4) in the S-SD structure at $z = L$.

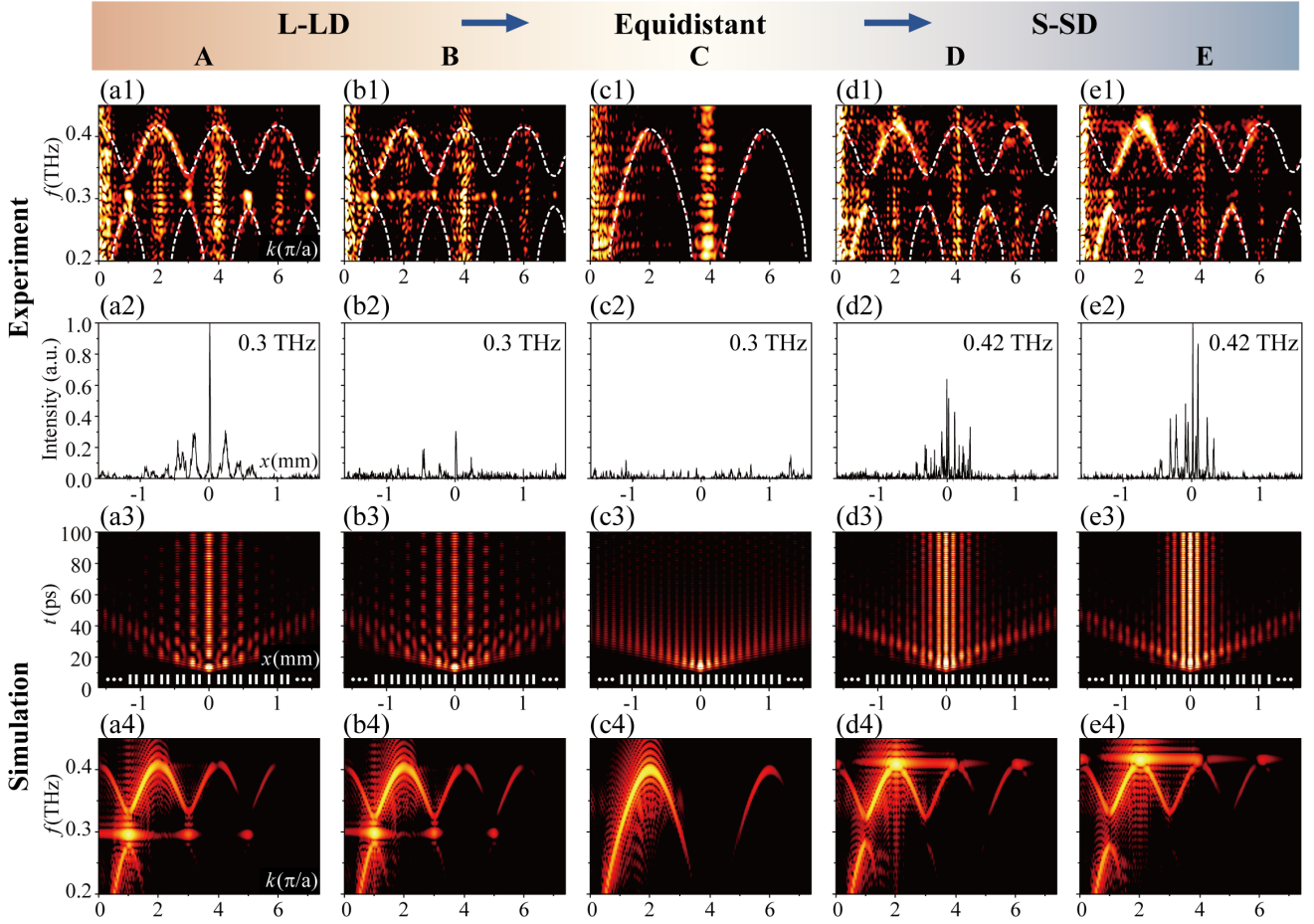

Fig. 3. Experimental (top two rows) and numerical (bottom two rows) demonstrations of topologically controlled THz confinement in the LN chip from L-LD to equidistance, and then to S-SD regions of the wedge-shaped SSH photonic lattice. A, B, C, D and E correspond to locations marked in Fig. 1(b). (a1-e1) Measured spectra at corresponding positions. (a2-e2) Energy distribution of the modes showing different confinement of the generated THz waves in the LN chip. (a3-e3) Simulated $x$-$t$ diagrams showing evolution of the THz waves in different regions, and (a4-e4) are the corresponding spectra. The lattice sites are illustrated by white tick marks in (a3-e3), and a is the lattice constant of the L-LD structure.

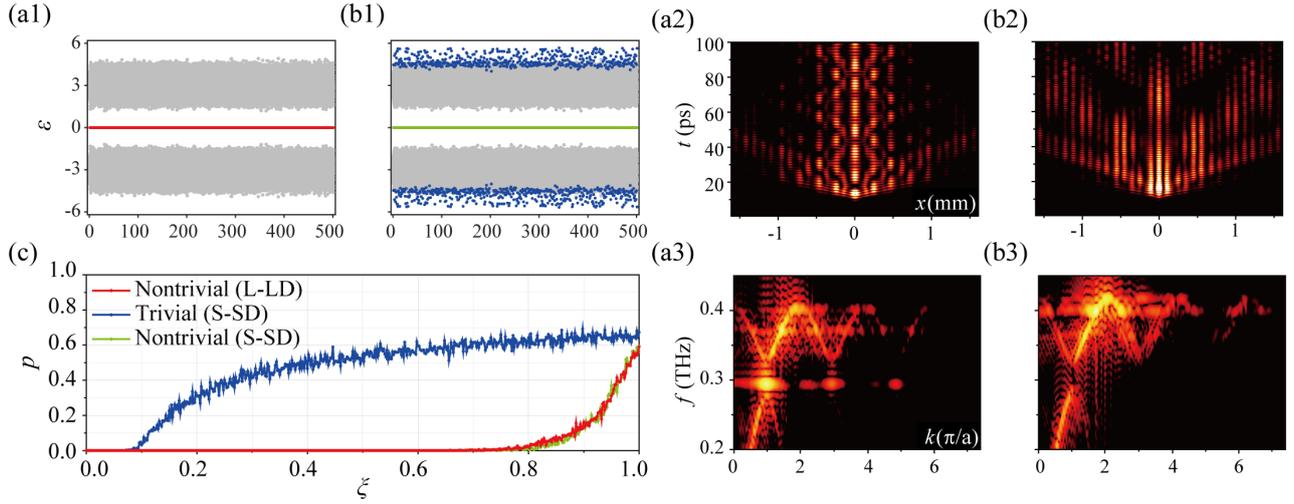

Fig. 4. Distinction between topological nontrivial and trivial defect modes under chiral perturbations. (a1) Theoretical calculation of eigenvalue distribution of L-LD structures under 500 sets of off-diagonal perturbations with $\max(\xi_n) = 30\%$. The red dots represent the eigenvalues of topological mode and gray dots show the distribution of bulk modes. (a2) Simulation of *x-t* diagram for the center defect excitation under perturbations. (a3) The corresponding spectrum of (a2). (b1-b3) have the same layout as (a1-a3) but for S-SD structures, where green and blue dots denote topological nontrivial and trivial defect modes, respectively. (c) The probability of defect-bulk mode coupling $p$ versus perturbation strength $\max(\xi_n)$. Red and green lines illustrate the nontrivial modes in L-LD and S-SD structures, respectively, and blue line is for the trivial defect mode in S-SD structures.